\documentstyle[12pt]{article}
\begin{document}

\title{\bf On gamma-ray bursts and their biological effects : 
a case for an extrinsic trigger of the Cambrian explosion ?}

\author{J. E. Horvath\\
\it Instituto de Astronomia, Geof\'\i sica e
Ci\^encias Atmosf\'ericas,\\  Universidade de S\~ao Paulo,\\
R. do Mat\~ao 1226 (05508-900) S\~ao Paulo SP, Brazil\\}

\maketitle

\vskip5mm

\begin{abstract}
We discuss some the effects of local gamma-ray bursts on the
earth's atmosphere. A rough calculation of the fraction of ozone
destruction by catalytic $NO_{x}$ cycles is given, which in turn
serves to argue how the large flux of gammas from these events
would have indirectly provoked major extinction of living
organisms. We give specific examples of these features, and
tentatively identify the Cambrian explosion seen in the actual
fossil record as an event caused by a GRB.
\end{abstract}

\clearpage

\section{Introduction}

A breakthrough in GRB astrophysics has been achieved by the
observation of afterglows located $\sim $ few hours after the
event with unprecedented positional accuracy (\cite{cos97}). The
presence of absorption lines (\cite{met97,kul98}) in some of these
afterglows has convinced most researchers that most (if not all)
of the GRBs are extragalactic, although a through comprehension of
the bursts is still far away since the sources have yet to be
identified and the physics of the afterglows addressed (see, for
example, \cite{ree97,hur98}). Nevertheless, we may now assert that
a distance scale (and hence an energy scale) is available for
"classical" bursts. Regardless of the specific source, it is now
clear that the evidence points out to $E_{\gamma}$ as high as
$\simeq \, 10^{53} \, erg$ for the most energetic bursts
(\cite{kuk98}) if the gamma emission is isotropic. Given that such
an extreme energy is quite difficult to obtain, and some
convincing observational features have accumulated, the idea of a
strongly beamed gamma flux has gained acceptance. A "standard
reservoir" of $\sim 10^{51} erg$ has been advocated by Frail et
al. \cite{ku} in a recent analysis.

While the study of distant, frequent bursts continues, the
observations have undoubtedly risen a number of questions related
to the occurrence of GRBs in the {\it local} universe. Thorsett
\cite{tho95} has discussed the effects of a close GRB on the
earth's biosphere (see also \cite{DLS98} for a related
discussion). The issue is timely since it has been shown that a
burst must occur as often as ($0.3-40$) $Myr$ per $L_{\ast}$,
depending on the evolution of the sources (\cite{wij98}). Loeb and
Perna \cite{LP} have further suggested that most of the HI
supershells could be the remnants of GRBs. In fact some remnants
must be found in a given normal galaxy, since they should not
dissipate before $\sim$ tens of $Myr$. The two gigantic shells
reported in \cite{RV93} in NGC 4631 are perhaps the most clear
examples that $\sim$ kpc-sized shells requiring $\sim \, 10^{54}
\, erg$ of input energy are real since their identification is
neater in external galaxies. The incidence of gamma rays from a
GRB should be then seriously considered.

\section{Gamma ray fluxes onto the earth and the ozone layer}

Consider the case of the simplest, "standard candle" scenario for
GRBs. We may estimate immediately the flux of gamma-rays at the
typical band 30-2000 keV at the top of the atmosphere $\phi$ for
an assumed energy in gamma-rays of $E_{\gamma} \, = \, 10^{51} \,
erg$. Since the true luminosity distribution function is still an
unsettled question and there might be a considerable spread
between the events, other possibilities should be considered.

Thorsett \cite{tho95} pointed out that GRBs this close would
(because of the huge gamma fluxes) should have produced deep
effects on the biosphere. The destruction of a substantial amount
of the ozone layer along a $\sim \, 10 \, s$ typical burst
duration is the most obvious one, and seems inescapable since the
$\geq \, 10^{7} erg \, cm^{-2}$ gamma fluxes  are in fact larger
than the equivalent total chemical energy of the fragile ozone
layer.

As discussed by Schramm and Ellis \cite{SE95} (see \cite{rud74}
for the first discussion of a closely related event), several
general features of the incidence of a huge gamma flux can be
worked out with some confidence. For example, it is well
established that the production of large concentrations of odd
nitrogen $NO_{x}$ is very harmful for the fragile ozone layer
shielding the earth from solar UV radiation. The dominating
catalytic reactions are

\begin{equation}
NO \, + \, O_{3} \, \rightarrow \, NO_{2} \, + O_{2}
\end{equation}
\begin{equation}
NO_{2} \, + \, O \, \rightarrow \, NO \, + \, O_{3} \;,
\end{equation}

since their efficiency of ozone destruction is high. The
additional $NO$ produced by the ionizing gamma flux will greatly
enhance the penetration of solar UV because the former is expected
to be much higher than the steady production by normal cosmic
rays. The rate of production of $NO$ (in $mol/cm^{2}$) is

\begin{equation}
\xi \, = \, 10^{17} \phi_{7}
{\biggl[ {13\over{10 + y}}\biggr]} \;,
\end{equation}

where $\phi_{7} \equiv (\phi/10^{7} erg \, cm^{-2})$ is the
incident gamma flux scaled to a reference value, and the factor in
brackets is the ratio of efficiencies of the steady production to
the GRB flash in the stratosphere. Dividing $\xi$ by the
stratospheric column density and converting to parts per billion,
we derive the abundance of $NO$ produced by the GRB flash as the
physical solution of a quadratic equation, very well approximated
by

\begin{equation}
y_{flash} \, \simeq \, 51 \phi_{7}^{1/2} \, - \, 5 .
\end{equation}

Thus, the ratio of produced $[NO]$ to the present ambient
$[NO]_{0}$ is given by $X = (3+y_{flash})/3 \, \sim \, 16
\phi_{7}^{1/2}$. Such a great abundance of $NO$ would remain in
the stratosphere for a mean residence time of $<\tau> \, = \, 4 \,
yr$, which is much larger than the homogenization time of the
atmosphere. Thus, once produced by the flash the ozone layer would
be affected for a period at least as large as the mean residence
time of the catalyzer.

The approximate formula employed in \cite{SE95} and \cite{rud74}
to estimate that reduction is

\begin{equation}
{{[O_{3}]\over{[O_{3}]_{0}}}} \, = \,
{{(16 + 9 X^{2})^{1/2} - 3 X}\over{2}} \;,
\end{equation}

expected to be accurate to within a numerical factor. It must be
noticed that, according to the Ruderman-Schramm-Ellis results, the
ozone destruction curve rises very rapidly with the gamma flux,
presenting a "catastrophic" destruction  which kills at least $90
\%$ of the present $O_{3}$ layer through $NO$ enhancement for a
fluence $\phi \geq \, 0.7 \times 10^{7} \, erg cm^{-2}$. Actually it
is highly likely that $\sim$ tens percent $O_{3}$ destruction
would already trigger massive biological death. Recent work
by Gherels et al. \cite{ghe02} using a detailed radiative transport
code arrives to a much lower figure of $\sim \, 10^{8} \, erg cm^{-2}$
for a $\sim \, 35 \%$ ozone depletion. Nevertheless, the
important point to stress is that a large flux like this figure is
actually expected from a galactic beamed burst which pointed to
the earth if that ever happened.

In order to estimate some of the effects onto the biota we shall
begin by calculating, using a simple attenuation model, the
killing timescale of a marine unicellular organism population
exposed to the UVB (260-320 $nm$) radiation immediately after the
burst. After a huge reduction of $O_{3}$ like the one discussed
above one may assume the solar flux density to be essentially the
value measured at the top of the atmosphere, some $0.2 \, W \,
cm^{-2} \, \mu m$ on average. The mortality of single-cell
organisms by UBV photons can be described (\cite{photo}) by

\begin{equation}
{N \over{N_{0}}} \, = \, 1 - {\bigl( 1 - \exp(- \kappa D)\bigr)}^{m}
\end{equation}

where $N_{0}$ is the original population, $m$ is the number of
absorbed photons needed to kill the cell, $\kappa$ is a constant
depending of the species and $D$ is the dose here defined as $\int
F_{\lambda} d\lambda$. This model can be immediately applied to a
marine population assumed to be distributed exponentially with a
fixed depth scale $z_{0}$ (i.e. without considering day-night
circulation) and having a spontaneous reproduction rate $\eta$. If
$N_{s}$ is the number of organisms at $z = z_{0}$ and the
coefficient of attenuation for UVB photons in marine water is
$z_{1}$ we find that the temporal evolution of this population at
any depth will be given by

\begin{equation}
N(z, t) = N_{s} \exp(-z/z_{0}) \, \exp[(\eta - \xi(z))t]
\end{equation}

with $\xi(z) \, \simeq \, \kappa F_{\lambda 0} \, \Delta \lambda
\exp(-z/z_{1})$. Now we may ask which is the time for killing $90
\%$ of these organisms once the UVB flux starts to impact onto the
sea surface, denoted as $\tau_{90}$. If we normalize the mortality
curve using the data from modern bacteria (i.e. {\it Escherichia
Coli}), for which plenty of data is available, we obtain for this
time $\tau_{90} = 0.4 \exp(z/z_{1}) \, s$. Therefore, it is
concluded that simple marine organisms, and especially those
capable of photosynthesis, will be killed almost instantaneously
unless they can "hide away" at several tens of $z_{1}$, in
practice $\geq 100 \, m$ for a time as long as the healing of the
ozone layer, which is certainly larger than most of the small
marine organisms considered. Terrestrial organism behavior is much
more difficult to model, although it has been known for fifty
years that mammals would not survive longer than $\sim \, 1 s$
without ozone. Even though simple models may be oversimplified
(they ignore all the detailed DNA repairing mechanisms and assume
an unimpeded single value of the solar UV flux following the
event), we believe that the essential points of a mass extinctions
by a GRB are adequately illustrated and quantified beyond any
reasonable doubt.

The gamma shower would have produced other rather unique
catastrophes as well, all them contributing to the extinction of
living beings. The production of $\sim \, 10^{9} \, tons$
of $NO_{x}$ enhancing the acid rains and the screening effects of
$NO_{2}$ to the sunlight (with possible dramatic cooling effects,
see \cite{rei78}) are just two of them. To address these issues
properly, the actual possibility of a close GRB calls for a
through study of the dynamical response of the biosphere to a
large perturbation, since all the effects are deeply interwoven
and it is quite difficult to isolate them due to their non-linear
character.

\section{A tentative association of a GRB with the Cambrian explosion}

According to the picture above, it seems quite clear that the
incidence of a gamma-ray burst beam onto the atmosphere would
trigger a quite remarkable biological evolution pattern, yet to be
precisely characterized. The natural question is whether we may
associate a definite event to such an external cause, based on the
existing evidence and a bit of extrapolation. As discussed above,
the extinction of a large number of living species due to UVB
action is, of course, a first necessary feature, but there are a
few more related effects to consider. One is the fate of the {\it
surviving} populations, since they are likely to be exposed to a
large ultraviolet flux on average, a powerful force driving their
further evolution through its action on the genetic material
(hypermutation?). It is also important to have in mind that the
surviving populations may be physically isolated from each other
in relatively small ecosystems, a scenario that can be justified
by the very nature of the catastrophic event. This situation is
known to be favorable for rapid speciation, although it is not as
certain whether a hypermutation rate drives an accelerated
evolution \cite{AF00}. The exposure of bacteria to UV light is
known to have dramatic effects by exploding the cell and releasing
bacteriophage genes, another feature that could have suddenly
boosted the genetic exchange and cleared the ecosystem from very
abundant, well-adapted organisms at the same time. 

We suggest that the celebrated Cambrian explosion, $\sim 544$ Myr
ago may grossly match these features. It is now established that
the oxygen levels were high enough at that epoch, and it is fair
to assume an ozone layer essentially equal to the modern one. At
the Proterozoic-Cambrian boundary and before the explosion itself,
simple animals pertaining to the diploblastic Ediacaran fauna were
suddenly extinct on a very short timescale; and the emergence of
an extremely diversified fauna followed, suggesting some global
event that could have triggered genetic experiments eventually
leading to an exponential growth of the number of species during
the early Cambrian age. While several intrinsic causes have been
advanced to justify this rising (see \cite{KC} for a review), it
is intriguing to consider the rather strong evidence for the
explosion being analogous to the case of the K/T boundary. In the
former, simple Ediacaran organisms played the role of "dinosaurs",
as emphasized by Knoll and Carroll (\cite{KC}), then followed by
an enormous diversification of the species. Contrary to the K/T
case, however, the postulated external trigger (illumination by a
GRB) would not leave an obvious signature like the famous iridium
layer, and thus subtle evidence should be searched for carefully.
In particular, the large and short-lived negative excursion in the
carbon-isotopic composition of surface seawater (also present but
with a smaller amplitude in the Permo-Triassic extinction) may
hold the clue for understanding the likelihood of a large
environmental perturbation. An explicit modelling of the dynamics
of the populations subject to such a large perturbation (just
underway) would shed some light onto this (yet speculative)
association.

\section{Conclusions}

We have shown that a local GRB gamma flux onto the earth should
correlate with a massive extinctions of life in the past. We are
still exploring the possible role of GRBs as punctuating agents of
the biological evolution, and the purpose of this paper has been
to show and quantify some of these effects. As discussed in the
previous section, we believe that the Cambrian explosion could be
a major example of the GRB role. Arguments for extrinsic
perturbation associated with an extinction event have been also
presented by Ben\'\i tez, Ma\'\i z-Apellaniz and Canelles
\cite{gait} (a close supernova event) and Melott et al. \cite{pep}
(a tentative GRB-Ordovician extinction). An earlier version of the
present work, focused on the issue of a hypernova-supershell
connection has been released as a preprint \cite{eu}. It is
interesting (but perhaps not significative) to note that some HI
supershells (tentatively associated with GRBs as their putative
remnants \cite{LP}, like GSH 139-03-69 should have been almost
simultaneous with the Priabonian extinction around 35 Myr ago
where cool-temperature-intolerant organisms gradually died,
whereas GSH 242-03+37 has a characteristic age of 7.5 Myr where
even $^{10}Be$ marine sediments could be used for testing purposes
(\cite{mor91}). Independently of this, major extinction events
older than $\sim \, 10 Myr$ would not obviously correlate with
supershells or other similar structure which should be long
dissipated

As stated above, detailed calculations of the response of
atmospheres to gamma rays have been recently published
\cite{SSW03,ghe02}. In particular, the model of
Ruderman-Schramm-Ellis has been found to provide an overestimate
of the ozone destruction. This is just an example of the type of
uncertainties one encounters when dealing with this formidable
problem. Clearly, much work is needed before we pinpoint and
understand the nature and consequences of these events for the ISM
and biological activity with confidence.

\section{Acknowledgements}

We would like to acknowledge E. Reynoso for useful advice on HI
observations. We are also grateful to E. M. G. Dal Pino, G.
Lugones and G.A Medina Tanco for encouragement and discussions on
these subjects. This work has been supported by FAPESP  Foundation
(S\~ao Paulo, Brazil). Steward Observatory colleagues and staff
are acknowledged for creating a stimulating working atmosphere
during the early realization of this work.

\end{document}